\documentstyle[11pt,fullpage,here,hangcaption]{article}
\begin{document}

\def\lsim{\:\raisebox{-0.5ex}{$\stackrel{\textstyle<}{\sim}$}\:}
\def\gsim{\:\raisebox{-0.5ex}{$\stackrel{\textstyle>}{\sim}$}\:}

\begin{flushright}
TIFR/TH/94-35
\end{flushright}
\bigskip
\bigskip
\begin{center}
\Large{\bf SCENARIOS AND SIGNALS OF VERY HEAVY NEUTRINOS} \\
\end{center}
\bigskip
\bigskip
\hskip 4cm \vbox{\hbox{Probir Roy}
\hbox{Tata Institute of Fundamental Research}
\hbox{Homi Bhabha Road}
\hbox{Bombay 400 005, India}}

\bigskip
\bigskip

{\it Talk given at the International Conference on
Non-accelerator Particle Physics (Indian Institute of Astrophysics,
Bangalore, January, 1994 and at the BCSPIN Summer School,
Kathmandu, Nepal, March, 1994)}

\bigskip
\bigskip
\bigskip

{}~~~~~~~~~~$\bullet$ {\bf PRELIMINARY REMARKS} \\

{}~~~~~~~~~~$\bullet$ {\bf THEORETICAL SCENARIOS} \\

{}~~~~~~~~~~$\bullet$ {\bf COSMOLOGICAL CONSTRAINTS} \\

{}~~~~~~~~~~$\bullet$ {\bf COUPLINGS OF VERY HEAVY NEUTRINOS} \\

{}~~~~~~~~~~$\bullet$ {\bf PRODUCTION MECHANISMS AND DETECTION} \\

{}~~~~~~~~~~$\bullet$ {\bf SUMMARY} \\

\newpage

\noindent $\bullet$ {\bf PRELIMINARY REMARKS} \\
\bigskip

Neutrinos are very special elementary fermions.  Their uniqueness lies
in not having any electromagnetic charge.  This enables massive
neutrinos to be either Dirac fermions with distinct particles and
antiparticles or, giving up lepton conservation, self-conjugate
Majorana fermions.  (The distinction becomes meangingless in the
massless case when a Dirac fermion simply becomes a linear
superposition of two Majorana ones.)  For quite some time now there
have been speculations [1,2] on the existence of very heavy neutrinos.
This is the subject of my talk.

\bigskip

I shall discuss the interest in the properties of very heavy neutrinos
(generically denoted as $N$) lying in the mass range $$ 45 ~{\rm GeV}
< M_N < {\cal O}\; ({\rm TeV}).  $$ The lower bound is dictated by the
lack of observation at LEP 1 of the pair-production of any such
neutrinos.  The upper bound is flexible upto a factor of a few owing
to possible mixing angle uncertainties.  This bound stems from the
requirement [3] of perturbative unitarity in the neutrino
pair-production amplitude from the initial state of two longitudinally
polarized $W$ or $Z$ bosons.

\bigskip

It is convenient to consider alongwith $N$ the charge-conjugated field
$N^C \equiv (\bar N C)^T$ where $C$ is the charge-conjugation matrix
in spinor space.  For Dirac neutrinos $N \neq N^C$ whereas
Majorana ones obey the condition $N = N^C$.  The mass term for a heavy
Dirac neutrino can be written as
$$
{\cal L}^D_{mass} = -M^D_N \bar NN = -M^D_N (\bar N_L N_R + \bar N_R
N_L),
\eqno (1)
$$
where $N_{L,R} \equiv {1\over2} (1 \mp \gamma_5)N$.
There are four independent chiral components in this case: $N_L$,
$N_R$, $N_L^{~C} = N^C_{~R}$ and $N_R^{~C} = N^C_{~L}$.  In contrast,
a massive Majorana neutrino admits a term such as
$$
- {1\over2} M^M_N \bar N^C N + h.c. = - {1\over2} M^M_N (N_L^T C^{-1}
N_L + N^T_R C^{-1} N_R) + h.c.,
$$
where use has been made of the relation $\gamma^0 C^\star \gamma^0 =
C^{-1}$ in the last step. More generally, Lorentz invariance allows
separate Majorana masses $M^M_{L,R}$ for $N_{L,R}$ so that one can
take
$$
{\cal L}^M_{mass} = - {1 \over 2} M^M_L N^T_L C^{-1} N_L - {1 \over 2}
M^M_R N^T_R C^{-1} N_R + h.c.
\eqno (2)
$$
The most general mass term would be a sum of (1) and (2).

If $N_L$ transforms as a doublet and $N_R$ as a singlet of the
electroweak $SU(2) \times U(1)$ gauge group, $M^M_L (M^M_R)$ would
have to arise from the VEV of a Higgs field transforming as a triplet
(singlet).  Since a triplet representation usually has problems with a
unit value of the $\rho$-parameter, it is customary to take $M_L = 0$
with the caveat that high scale physics may induce a tiny $M_L$.  On
the other hand, a Higgs singlet -- being outside the electroweak gauge
theory -- could only be a relic of high scale physics so that one
expects $M_R = M \gg M_W$.  The general mass term for $N$ can then be
written in matrix form as $$ {\cal L}_{mass} = - {1\over2}
(\overline{N_L} ~\overline{N_R^{~C}})
\left(\matrix{0 & m_D \cr m_D & M}\right) \left(\matrix{N_L^{~C} \cr
N_R}\right) + h.c.
\eqno (3)
$$

In the seesaw [4] mass-matrix of (3) the off-diagonal element $m_D$ is
the Dirac mass.  It may be expected to be typically of the order of
the known charged fermion masses of the concerned family in the
Standard Model.  Of the two eigenvalues of the seesaw mass matrix, one
$(\sim M)$ is therefore expected to be heavy and the other $(\sim
m^2_D/M)$ light.  In this simplest of seesaws the physical particles
are Majorana fermions, but more complicated seesaws exist [5] where
they are of the Dirac type.  Furthermore, arguments have been given
[6] that the seesaw mechanism may be induced radiatively at the
electroweak 1-loop level.  Thus the seesaw formula for a light
neutrino might in fact be $m_\nu \sim \pi^{-1} \alpha_W m^2_D/M$,
where $\alpha_W$ is the weak fine structure constant.  In this case a
physical electron neutrino mass $\sim 10^{-2}$ eV (of interest to the
solar neutrino puzzle) and a Dirac mass ${\cal O}$~(MeV) would suggest a
heavy righthanded neutrino mass $M \sim 10^2 - 10^3$ GeV which may be
within reach of production and detection in the forthcoming colliders.
\bigskip

\noindent $\bullet$ {\bf THEORETICAL SCENARIOS} \\
\bigskip

There exist a number of scenarios in which such very heavy neutrinos are
expected to occur.  We outline a few.

\bigskip

\noindent \#1. {\sl Fourth generation model} -- This has been proposed by
Hill and Paschos [7] and has a heavy charged lepton $\ell$ and a
neutrino $N$ making a fourth replica of the existing three generations.
Thus one has a left-chiral doublet and two right-chiral singlets:
$$
\left(\matrix{N \cr \ell^-}\right)_L, ~~~\ell^-_R, ~~~N_R.
$$
Existing LEP constraints from $Z$-decay simply require that $M_N >
{1\over2} M_Z$.  One need not have $N_R$, but then one would be
forced to invoke a lepton-number violating Majorana mass term.  This
model has been shown to be natural in terms of flavor democracy [8]
which invokes a permutation symmetry in the $4 \times 4$ fermion
mass-matrix.

\bigskip

\noindent \#2. {\sl Left-right symmetric model} -- The simplest such model
employs the gauge group $SU(2)_L \times SU(2)_R \times U(1)_{B-L}$ and
contains one left-chiral and one right-chiral lepton doublet per generation:
$$
\left(\matrix{\nu \cr e}\right)_L,~~~\left(\matrix{N \cr e}\right)_R.
$$
These describe a very heavy
and a very light physical neutrino.  References to discussions of their
phenomenology can be found in [9].

\bigskip

The very heavy neutrino, described in the above two scenarios, can be
searched for in LEP 200 via the reaction $e^+e^- \longrightarrow
N\overline{N}$.  This process can proceed both by $Z$-exchange in the
$s$-channel and by
$W$-exchange in the $t$-channel, provided the $N$-mass is less than a 100
GeV.  For a higher mass, one would need to wait for the Next Linear
Collider (NLC) where picobarn
level cross sections are predicted for $\sqrt{s} \sim 300$ GeV.  It can, in
principle, be looked for by exploiting its mixing with $\nu_e$ via the
production mode $ep \rightarrow N + {\rm `anything'}$ at HERA, but the
estimated cross sections [10] look impossibly small.

\bigskip

\noindent \#3. {\sl Pure singlet model}  -- In this case there is an extra
right-chiral singlet heavy neutrino $N_R$ for each generation.  Thus, for
the first, one has
$$
\left(\matrix{\nu \cr e}\right)_L, ~e_R, ~N_R.
$$
$N_R$ can have a large Majorana mass and it is possible to arrange a
seesaw mass-matrix between $\nu_L$ and $N_R$.  The main importance of this
type of a model is that $N$ can be produced singly in $e^+e^-$ or
$ep$ collision, as detailed later.

\bigskip

\noindent \#4. {\sl $E_6$-based models} -- The grand unifying gauge group
$E_6$ is very popular among model builders starting with the $E_8 \times
E_8$ superstring.  The matter fields in an $E_6$ GUT, arising from the
topological breakdown of one of the $E_8$'s, belong to the
{\bf 27} dimensional representation of $E_6$.  This multiplet
accommodates three extra heavy neutrinos [11], their masses depending on
various symmetry-breaking scales in the breakdown chain $E_6 \rightarrow
SU(3)_C \times SU(2)_L \times U(1)_Y$.  Two of these neutrinos are
singlets under $SU(2)_L \times U(1)_Y$.  The third, transforming as part of a
vectorial doublet with respect to weak isospin, directly couples to $W$
and $Z$.  The others too can mix with the usual very light neutrinos
$(\nu_\ell,\ell = e,\mu,\tau)$ and develop couplings to the weak bosons.

\bigskip

\noindent \#5. {\sl Supersymmetric preon models} -- These [12] also
contain the very heavy neutrinos of the kind discussed in $E_6$-based
scenarios.  Additional ones may be possible but the detailed
phenomenology has not been worked out.

\bigskip

\noindent $\bullet$ {\bf COSMOLOGICAL CONSTRAINTS} \\
\bigskip

We shall discuss these under three headings.

\bigskip

\noindent \#1. {\sl Stable very heavy neutinos} -- The discussion of
constraints on the properties of very heavy neutrinos from early
universe considerations depends on whether those neutrinos are stable
or unstable.  In the former situation the universe today would be
pervaded by a sea of such objects.  We shall keep in mind two generic
cases: (1) a stable lefthanded $N$ transforming like an $SU(2)_L$
doublet [7] with a universal four fermion coupling characterized by
the Fermi constant $G_F$; (2) a stable nonstandard $SU(2)_L$-singlet
$N$ (say righthanded), with a four-fermion coupling of at least
milliweak strength $(\gsim 10^{-3} ~G_F)$ which is induced by some
kind of mixing.

There is an upper bound [13] on the sum of the masses of stable
standard neutrinos from the requirement of not over-closing the
universe through excess mass-density.  However, the commonly written
form, i.e.
$$
\sum_\nu m_\nu < 100 h^2\; eV,
$$
where $h \equiv ({\rm Hubble~constant}/100~ km~s^{-1}~ Mpc^{-1})$, is
inapplicable to $N$ if its mass exceeds 2 GeV.   Such heavy
neutrinos equilibriate in the early universe by pair-creation and
annihilation.  These processes continue to occur till the universe
cools below their freeze-out temperature $T_f$ at which point they go
out of chemical equilibrium.  The argument for the above inequality
then needs to be reformulated in terms of statistical mechanics and
yields [14]
$$
M_N e^{-M_N/(kT_f)} < 100 h^2\; eV,
\eqno (4)
$$
$k$ being the Boltzmann constant.  The freeze out temperature for very
heavy neutrinos has been estimated to be [15] $T_f \sim M_N/(20 k)$.
Thus (4) is easily satisfied for a Dirac mass in excess of $2$ GeV.

\begin{figure}[H]
\vspace{6in}
\hangcaption{Cosmological constraints on stable heavy neutrino masses.}
\end{figure}

Even in the post-freeze out era, however, the reaction $N\overline{N}
\rightarrow e^+e^-$ can continue to take place.  Subsequently, the
produced $e^+e^-$ pair can annihilate into two photons.  If
unrestricted, this process will generate far too many photons and lead
to an unacceptably high value of the entropy density of the universe.
The necessary restriction is
$$
M_N (T/T_\gamma)^3 < 100h^2~ eV,
\eqno (5)
$$ where $T_\gamma$ is the temperature of the cosmic microwave
background radiation while $T$ is the temperature that would have
accrued to the sea of $N$'s, if $N$ were massless.  When (5) is
applied to case (1), the restriction $M_N < 3$ TeV follows [16],
while, for case (2), the corresponding result is $M_N < 10$ TeV for
$G/G_F \sim 10^{-3}$.  Studying the latter case in more detail [17],
Olive and Turner have obtained quantitative restrictions on $M_N$ as a
function of $G_F/G$ as shown in Fig. 1 for Dirac and Majorana $N$'s.

Additional constraints on the masses of very heavy stable neutrinos
have emerged [18] from dark matter search experiments.  The
annihilation of $N\overline{N}$ into charged particles is exploited
for this purpose by seeking to detect the latter.  The restrictions,
consequent upon the lack of such detection, depend on the assumptions
made in calculating the flux of $N$'s.  There are two possible
approaches.  (a) If one assumes that the galactic halo is composed
solely of such stable very heavy neutrinos, the constraints obtained
on their masses, for the Dirac and Majorana cases respectively, are:
$$
\begin{array}{l}
M^D_N < 6 ~{\rm GeV~or}~ M^D_N > 300 ~{\rm GeV}, \\[2mm]
M^M_N < 24 ~{\rm GeV~or}~ M^M_N > 300 ~{\rm GeV}.
\end{array}
$$
(b) On the other hand, if the flux is calculated by considering the
relic abundance of $N$'s taking such annihilation processes as
$N\overline{N} \rightarrow f\overline{f},W^+W^-,ZZ$ ($f$ any quark or
lepton) into account, the corresponding results become
$$
\begin{array}{l}
M^D_N < 6 ~{\rm GeV~or}~ 40 ~{\rm GeV}~ < M^D_N < 48 ~{\rm GeV},
\\[2mm] M^M_N < 24 ~{\rm GeV~or}~ 38 ~{\rm GeV}~ < M^M_N < 52 ~{\rm
GeV}.
\end{array}
$$

\bigskip

\noindent \#2.  {\sl Unstable very heavy neutrinos} -- The possible
channels for the decay of an unstable $N$, that have been considered,
are : $N \rightarrow \nu \gamma$ and $N \rightarrow \nu \ell^+
\ell^-$, $\ell$ being a charged lepton.  If $N$ is very massive and
decays late in the evolutionary history of the universe, the decay
products will make a significant contribution to the energy density of
the universe, albeit redshifted by the Hubble expansion.  This can
yield any meaningful constraint [19] only if the lifetime of $N$ is
greater than a second or so.  For a highly unstable $N$, which is of
interest to accelerator-based particle physicists, there are
practically no constraints from cosmology.

\bigskip

\noindent \#3. {\sl Leptogenesis} -- The existence of a massive
Majorana neutrino $N$ is perforce in contradiction with lepton
conservation since the mass term acts as a lepton number violating
operator.  Thus the presence of such particles would mean a lepton
asymmetric early universe in which leptogenesis occurred at the (GUT)
timescales when $N$ acquired its mass.  However, during the
electroweak phase transition, sphaleron-induced baryon and lepton
nonconserving (but $B-L$ preserving) processes at temperatures between
about 100 GeV and 1 TeV would convert this leptonic asymmetry into a
baryonic one.  This is a viable [20] scenario (i.e. the sphaleron
processes can remain in thermal equilibrium for the requisite period)
for a whole range of heavy neutrino masses from about 1 TeV to
$10^{12}$ GeV.  It has also been demonstrated [21] that, even if the
conservation of the full $B-L$ is invalid, that of ${1\over3} B-L_i$ --
where $i$ refers to {\sl any} leptonic type -- is sufficient for the
mechanism to go through.
\bigskip

\noindent $\bullet$ {\bf COUPLINGS OF VERY HEAVY NEUTRINOS} \\

\bigskip

There are far too many model-independent possibilities in the pattern of
couplings of such very heavy neutrinos to the known elementary particles.
We try to adopt a generic approach following [22].  Let us assume that, on
account of mixing with $\nu_\ell$, $N$ develops a charged current coupling
to $W\ell$ and neutral current couplings to $NZ$ as well as $\nu_\ell Z$
-- as shown in Fig. 2.  Here  $\xi$ is a small

\begin{figure}[H]
\vspace{3in}
\hangcaption{Heavy neutrino couplings.}
\end{figure}

\noindent seesaw mixing factor (hopefully $\gsim
10^{-3}$) and $V_{N\ell}$ is a Kobayashi-Maskawa type matrix element.
(It may be noted that the model of [7] cannot be covered by this since
there is no $ZN\bar\nu_\ell$ vertex there and no mixing factor in the
$ZN\bar N$ coupling).  We are also obliged to choose $M_N > M_Z$,
otherwise -- for $\xi > 10^{-3}$ -- the decays $Z \rightarrow
M\bar\nu_\ell,\bar N\nu_\ell$ would have already been seen at LEP.

We come to the decays of $N$.  First, consider the case when $N$ is a
Dirac particle.  Now, for the charged current mode
$$
\Gamma (N \rightarrow \ell^+ W^+) = \Gamma(\bar N \rightarrow \ell^+ W^-)
= {|\xi V_{N\ell}|^2 \over 8\sqrt{2} \pi} {G_F \over M^3_N} (M^2_N +
2M^2_W) (M^2_N - M^2_W)^2.
\eqno (6)
$$
Contrariwise, for the neutral current mode
$$
\Gamma(N \rightarrow \nu_\ell Z) = \Gamma(\bar N \rightarrow \bar\nu_\ell
Z) = {|\xi|^2 \over 8\sqrt{2}\pi} {G_F \over M^3_N} (M^2_N + 2M^2_Z)
(M^2_N - M^2_Z)^2.
\eqno (7)
$$ The equality of the $N$ and $\bar N$ partial widths in (6) and (7)
follows from $CP$-invariance.  The charged current mode is less
dominant than the neutral current one since the latter does not have
the small $|V_{N\ell}|^2$ factor.  For a Majorana heavy neutrino, $N$
and its antiparticle are identical.  Now one simply has $\Gamma(N
\rightarrow \ell^- W^+) = \Gamma(N \rightarrow
\ell^+ W^-)$ and $\Gamma(N \rightarrow \nu_\ell Z) = \Gamma(N \rightarrow
\bar\nu_\ell Z)$, with the corresponding expressions still given by (6)
and (7) respectively.  Hence the lifetime of $N$ gets halved as compared
with a Dirac N.  In either case, for $M_N \gg M_{W,Z}$ and $\xi \gsim
10^{-3}$, the mean free path is $\ll$ 1 cm.  Thus, if produced in the
laboratory, such an $N$ will decay within the detector.

Though the neutral current induced decay is the dominant mode, the charged
current mediated one $(N \rightarrow \ell W)$ can provide the cleanest
signals for detection.  The $W$ can decay into two jets so that $\ell(2j)$
is the detectable final state configuration.  For the Dirac case and with
a pair-produced $N\bar N$, one would have the hard signal $\ell^+
e^{^{\prime \atop -}} (4j)$ where $\ell$ and $\ell^\prime$ need not be
the same.  (Of course, one would have to tackle the severe background from
the semileptonic decays of top-antitop pairs).  If a pair of Majorana $N$'s
gets prodeuced, we can have three possibilities : $\ell^+ e^{^{\prime
\atop -}} (4j), ~\ell^+ e^{^{\prime \atop +}} (4j)$ and $\ell^-
e^{^{\prime \atop -}} (4j)$.  While these are characteristic signals,
one cannot exclude a very heavy neutrino in the relevant mass-range simply
by failing to observe them.  This is because certain models allow [23] the
dominant decay $N \rightarrow \nu J$ where $J$ is a very light
pseudo-Goldstone boson like a Majoron.  The experimental unobservability of
this decay channel would make it harder to discover $N$.

\bigskip

\noindent $\bullet$ {\bf PRODUCTION MECHANISMS} \\

\bigskip
\nobreak
First, we take up the production of single $N$'s.  In an $e^+e^-$
collider this can be done through the processes $e^+e^- \rightarrow
N\bar\nu_\ell,\bar N\nu_\ell$.  These go via the Feynman diagrams of
Fig. 3.  Asymptotically, for a large $CM$ energy $\sqrt{s}$, the cross
section is approximately $\pi^{-1} G^2_F M^2_W |V_{\ell N} \xi|^2$.
\begin{figure}[H]
\vspace{3in}
\hangcaption{Diagrams for single $N$ or $\bar N$ production in $e^+e^-$
collision}
\end{figure}
\noindent
Note that, for a Majorana $N$, there are only three diagrams since (c)
and (d) become one and the same.  In any event, the $Z$-mediated part
contributes only about 2\% of the total cross section.  Thus it is a
good approximation to take only the $W$-mediated part.  Typically, a
fraction of a picobarn is expected at LEP 200 as shown in Fig. 4, where
the production cross section [22] has been plotted against
$\sqrt{s}$ as well as against the heavy neutrino mass $M_N$.  Coming
to electroproduction $e^-p \rightarrow NX$, say at HERA, the cross
section is shown against $M_N$ for various values of $\sqrt{s}$.  Of
course, the signal (Fig. 5) will depend on whether the $N$ decays into
$\ell W$ or $\nu_\ell Z$, but $\ell^- (2j)\; /\!\!\!\!E_T$,
$\ell^+\ell^-\; /\!\!\!\!E_T$, and $\ell^+e^{^{\prime \atop -}}
/\!\!\!\!E_T$ are possible signal configurations.
\begin{figure}[H]
\vspace{3.7in}
\hangcaption{\protect
Cross sections for $e^+e^- \rightarrow N\bar \nu_e$ at LEP
200 as functions of $M_N$ and $\sqrt {S}$.}
\end{figure}
\begin{figure}[H]
\vspace{3.7in}
\hangcaption{Cross sections for $\bar e p \rightarrow NX$ at HERA as a
function of $M_N$.}
\end{figure}

Next, we come to pair-production.  This can be attained through the $ZN\bar
N$ coupling which is perhaps less model-dependent than the $ZN\bar \nu$
one.  We put a generic mixing factor $\chi$ to cover the cases where $N$
is an $SU(2)_L \times U(1)_Y$ singlet.  (For a regular fourth generation
heavy neutrino, $\chi$ is unity).  The cross section for $e^+e^-
\rightarrow Z^\star \rightarrow N\bar N$ (Fig. 6) can be calculated [11] to be
$$
\sigma = {G_F^2 s \over 24\pi} \left(1 - {4M^2_N \over s}\right)^{1/2}
\left(1 - {M^2_N \over s}\right) \left({M^2_Z \over M^2_Z + s}\right)
|\chi|^2 (1 - 4x_W + 8x^2_W).
\eqno (8)
$$
\begin{figure}[H]
\vspace{2in}
\hangcaption{Lowest order diagram for $e^+e^- \rightarrow N\bar N$}
\end{figure}

In the high-energy limit when $s \gg M^2_Z$, the RHS of (8) goes as $2.5
\times 10^{-2} |\chi|^2$ (pb/s in TeV$^2$) which is about $0.6 |\chi|^2$
pb for $\sqrt{s} = 200$ GeV at LEP 200.  Similar considerations hold for
the Drell-Yan type of production process $q\bar q \rightarrow Z^\star
\rightarrow N\bar N$ in a hadron collider.  One problem with the cross
section of (8) is the rapid fall off at large $s$
which drastically reduces the cross section at supercollider energies
$\gsim 1$ TeV.

We shall discuss an alternative mechanism of heavy neutrino
pair-production via the fusion of two gluons [24,25] which is relevant
to $pp$ supercolliders.  As shown in Fig. 7, two gluons from the
colliding protons can go via a quark loop into an off-shell $Z$ or
Higgs boson which converts into an $N\bar N$ pair.  The heavy
neutrinos, in turn, decay into $\ell (2j)$ and $\ell'(2j)$ final
states, say, so that the signal configuration is $\ell\ell' (4j)$.
For a Majorana pair, one can have like sign dileptons which with four
jets make an almost unique signal for this process.  In the case of a
Dirac pair and a signal configuration of $\ell^+e^{^{\prime \atop -}}
(4j)$, the background from $t\bar t$ pair-production and subsequent
semileptonic decays of $t,\bar t$ would be overwhelming.  But now one
can hook on to the leptonic decay of one of the $W$'s and search for
the signal $\ell^+ e^{^{\prime \atop -}} e^{^{\prime\prime \atop +}}
(2j)\: /\!\!\!\!E_T$ or $\ell^+ e^{^{\prime \atop -}}
e^{^{\prime\prime
\atop -}} (2j)\: /\!\!\!\!E_T$, where  $/\!\!\!\!E_T$ is the missing
transverse energy.
\begin{figure}[H]
\vspace{3in}
\hangcaption{Gluon fusion into decaying heavy neutrino pairs in $pp$ collision}
\end{figure}

There are several characteristic features of the $Z$-exchange mechanism:

\bigskip

$\bullet$ The triangular loop has a nonzero contribution only from the
axial part of the $Z$-coupling, the vector part vanishing on account of
Furry's theorem.

$\bullet$ The contributing part is an anomaly graph proportional, not only
to the third component of the weak isospin of the fermion circulating in
the triangle, but also -- through the divergence of the axial current --
to its mass.  Thus the mass difference $|M_U - M_D|$ between the up-type
and down-type quarks of the heaviest generation comes in the numerator of
the dominant part of the amplitude.

$\bullet$ By Yang's theorem, the amplitude is nonzero only because of the
off-shell nature of the $Z$.  The consequent $Q^2 - M^2_Z$ in the
numerator cancels the denominator from the propagator making the cross
section only weakly dependent on $s$.

In the light of the above features it turns out that the
Higgs-mediated $gg \rightarrow H^\star \rightarrow N\overline{N}$
cross section is enhanced relative to the $Z$-mediated one to which it
adds incoherently on-account of the difference in the $s$-channel
angular momentum.  One can actually have both scalar Higgs $H$ and
pseudoscalar Higgs $P$ exchanges (in models with more than one
doublets).  It has been [27] demonstrated
(taking $m_{\rm top} = 160$ GeV) that the cross section in either case
is expected to be quite large.  More recent works [28] have given
detailed discussions of such processes for the specific case when $N$
is a heavy righthanded Majorana neutrino.  In this case the discovery
limit for $N$ at the forthcoming Large Hadron Collider can go upto
$M_N \simeq 10~{\rm TeV}$.

\bigskip

\noindent $\bullet$ {\bf SUMMARY}

\bigskip

The salient features of current speculations on and searches for very
heavy neutrinos $N$ weighing more than 45 GeV can be summarized as
follows.

\begin{description}
\item[\rm (a)] One or more $N$'s of right chirality are needed to
implement the seesaw mechanism of generating the mass of a light
neutrino.  In case the mechanism is effected at the 1-loop level, such
$N$'s could have masses ${\cal O}\:(10^2~{\rm GeV})$ and be discovered
in the near future.

\item[\rm (b)] Several `reasonable' beyond-standard-model scenarios
exist predicting such objects.

\item[\rm (c)] If $N$ is stable or long-lived, its mass cannot exceed
a few TeV on account of cosmological constraints.  However, the mass
of a short-lived $N$ is essentially unconstrained by early universe
considerations.

\item[\rm (d)] Forthcoming collider experiments will seriously probe
mass regions $\sim 10^2~{\rm GeV}$ for an unstable $N$ decaying within
the detector.

\end{description}

\bigskip

\begin{center}
{\bf REFERENCES}
\end{center}

\begin{enumerate}
\item[{[1]}] M. Gronau, C.N. Leung and J.L. Rosner, Phys. Rev. {\bf
D29}, 2539 (1984).  H.E. Haber and M.R. Reno, Phys. Rev. {\bf D31},
2372 (1986).  F. Gilman, Comm. Nucl. Part. Phys. {\bf 5}, 231 (1986).

\item[{[2]}] P. Roy, in {\em Neutrino Physics} (Symposium on
Theoretical Physics, Mt. Sorak, Korea 1992, ed J.E. Kim, Min Eum Sa
Co., Seoul, 1993).

\item[{[3]}] J.D. Bjorken and C.H. Llewellyn Smith, Phys. Rev. {\bf
D7}, 887 (1973).

\item[{[4]}] T. Yanagida, Proc. {\em Workshop on Unified Theory and
Baryon Number in the Universe} (ed. O. Sawada and A. Sugamoto, KEK
1979).  M. Gell-Mann, P. Ramond and R. Slansky in {\em Supergravity}
(ed. P. van Nieuwenhuizen and D.Z. Freedman, North Holland, 1979).

\item[{[5]}] P. Roy and O. Shanker, Phys. Rev. Lett. {\bf 52}, 713
(1984); Phys. Rev. {\bf D30}, 1949 (1984).  M. Roncadelli and D. Wyler,
Phys. Lett. {\bf 133B}, 325 (1983).

\item[{[6]}] A. Pilaftsis, Z. Phys. {\bf C55}, 275 (1992).

\item[{[7]}] C.T. Hill and E. Paschos, Phys. Lett. {\bf B241}, 96
(1990).

\item[{[8]}] A. Datta and S. Raychaudhuri, Phys. Rev. {\bf D49}, 4762
(1994).

\item[{[9]}] G. Altarelli, M. Mele and R. R\"uckl, CERN-ECFA Workshop
{\bf 2}, 549 (1984) (QCD 183:E2:1984).

\item[{[10]}] J. Wudka, Phys. Rev. {\bf D49}, 1629 (1994).

\item[{[11]}] V. Barger and R.J.N. Phillips, {\em Collider Physics}
(Addison-Wesley, corrected edition, 1988).

\item[{[12]}] K. Babu, J.C. Pati and Stremnitzer, Phys. Lett. {\bf
B256}, 206 (1991); Phys. Rev. Lett. {\bf 67}, 1688) (1991).

\item[{[13]}] R. Cowsik and McCelland, Phys. Rev. Lett. {\bf 29}, 669
(1972).

\item[{[14]}] R.N. Mohapatra and P. Pal, {\em Massive neutrinos in
physics and astrophysics} (World Scientific, Singapore).

\item[{[15]}] B.W. Lee and S. Weinberg, Phys. Rev. Lett. {\bf 29}, 669
(1977).

\item[{[16]}] P. Hut, Phys. Lett. {\bf B69}, 85 (1977).  J.E. Gunn et al, Ap.
J. {\bf 223}, 1015 (1978).

\item[{[17]}] K. Olive and M. Turner, Phys. Rev. {\bf D25}, 213 (1982).

\item[{[18]}] M. Mori et al. Phys. Lett. {\bf B289}, 463 (1992).

\item[{[19]}] E. Kolb, Santa Cruz TASI (Theoretical Advanced Study
Institute,) 86:673.

\item[{[20]}] A. Dolgov, Nucl. Phys. {\bf B}, Proc. Suppl. {\bf 35},
28 (1994).

\item[{[21]}] H. Dreiner and G.G. Ross, Nucl. Phys. {\bf B410}, 188
(1993).

\item[{[22]}] W. Buchm\"uller and C. Greub, Phys. Lett. {\bf B256}, 465
(1991); Nucl. Phys. {\bf B381}, 109 (1992).

\item[{[23]}] Ref. 11, p459.

\item[{[24]}] D.A. Dicus and P. Roy, Phys. Rev. {\bf D44}, 1593
(1991).

\item [{[25]}] D. Choudhury, R.M. Godbole and P. Roy, Phys. Lett. {\bf
B308}, 394 (1993). errtm. {\em ibid}, {\bf B314}, 394 (1993).

\item[{[28]}] A. Datta, M. Gucahit and D.P. Roy, Phys. Rev. {\bf D47},
961 (1993) \\
A. Dutta, M. Gucahit and A. Pilaftsis, Phys. Rev. {\bf D50}, 3195 (1994).

\end{enumerate}

\end{document}